\documentclass[12pt]{article}
\usepackage{hepth}
\usepackage{graphicx}

\newcommand{\rhon}{{\rho_{\rm n}}}
\newcommand{\rhos}{{\rho_{\rm s}}}
\newcommand{\dualop}{\langle O_{\Delta} \rangle}

\begin{document}

\preprint{PUPT-2293}

\institution{PU}{Joseph Henry Laboratories, Princeton University, Princeton, NJ 08544, USA}

\title{Fourth sound of holographic superfluids}

\authors{Amos Yarom\worksat{\PU,}\footnote{e-mail: {\tt ayarom@princeton.edu} }}

\abstract{We compute fourth sound for superfluids dual to a charged scalar and a gauge field in an AdS${}_4$ background. For holographic superfluids with condensates that have a  large scaling dimension (greater than approximately two), we find that fourth sound approaches first sound at low temperatures. For condensates that a have a small scaling dimension it exhibits non-conformal behavior at low temperatures which may be tied to the non-conformal behavior of the order parameter of the superfluid.
We show that by introducing an appropriate scalar potential, conformal invariance can be enforced at low temperatures.}


\maketitle

\tableofcontents

\section{Introduction}
When ${}^4$He is cooled down to temperatures below $T_{\lambda} \sim 2.17\,^{\circ}$~K it undergoes a phase transition from a liquid phase, helium I, to a superfluid phase, helium II. While helium I behaves like a normal low-viscous fluid, helium II exhibits both viscous and non-viscous behavior. This property of helium II (and several other properties) can be explained by a two-fluid model first suggested by Tisza \cite{Tisza} and expanded upon by Landau \cite{Landau}. In this model, helium II is described as a fluid with two components. One, the normal component, has normal viscous behavior, while the other, the superfluid component, behaves like a frictionless fluid. 

Since the gauge-gravity duality \cite{Maldacena:1997re,Gubser:1998bc,Witten:1998qj} has been successful in describing the hydrodynamic regime of gauge theories with holographic duals \cite{Policastro:2001yc,Bhattacharyya:2008jc}, it is interesting to ask to what extent it can describe the hydrodynamic behavior of superfluids. 
To create a ``holographic superfluid,'' or a gravity dual of a system with a condensed phase, one needs a mechanism by which a condensate will be generated in an asymptotically AdS geometry. 
In \cite{Gubser:2008px} it was argued that a charged scalar coupled to an abelian gauge field in AdS space (an AdS version of the abelian Higgs model) may condense in the presence of a black hole horizon. From the point of view of the boundary theory  this corresponds to a second order phase transition whose order parameter is dual to the condensed scalar \cite{Hartnoll:2008vx}. 
Other types of symmetry breaking mechanisms were considered in \cite{Gubser:2008zu,Gubser:2008wv,Ammon:2008fc,Basu:2008bh}.

In \cite{Basu:2008st,Herzog:2008he} some similarities between the phase diagram of the condensate dual to the abelian Higgs model in the bulk, and that of a superfluid were discussed. Like a superfluid, the phase diagram of the boundary theory was shown to include a tricritical point and a critical superfluid velocity. A similar analysis was carried out in \cite{Basu:2008bh,Herzog:2009ci} for a holographic p-wave superfluid derived from a non-abelian gauge field in the bulk.

In this work we investigate the phase velocity of particular sound modes in holographic superfluids. In one component fluids, small perturbations of the pressure follow a wave equation with a phase velocity equal to the speed of sound. In superfluids, there are two propagating modes involving the pressure and the entropy density. As a consequence, superfluids support two sound modes and, appropriately, two phase velocities called first and second sound. In the limit where the density of the superfluid component goes to zero, second sound vanishes and the remaining sound mode is first sound---the one associated with pressure waves. Another limit can be obtained when an external force prevents the normal component from flowing. In this case a propagating mode involving density fluctuations will still remain. Since the normal component is stationary, these density fluctuations necessarily involve non-vanishing superfluid velocity. The phase velocity of such modes is called fourth sound \cite{Atkins}.\footnote{
Third sound, also defined in \cite{Atkins},
is the phase velocity of wave propagation on the surface of a thin film of superfluid.
}
See also \cite{PhysRev.73.608}. 
For ${}^4$He its value coincides with second sound near the phase transition, and first sound at low temperatures \cite{Atkins}. 

Going back to a holographic setup, in \cite{Herzog:2008he} the authors found an approximate expression for second sound, valid at high temperatures and the probe approximation (where the geometry does not back-react on the matter fields). In the following section we show that one may reinterpret the approximate expression obtained in \cite{Herzog:2008he}  as an exact expression for fourth sound, and then extend its range of validity to low temperatures, as long as the probe approximation holds. 

From the results of \cite{Herzog:2008he} one finds that, in the abelian Higgs model, at low temperatures, fourth sound squared approaches $1/2$ for dimension two condensates and $1/3$ for dimension one condensates  in $d=3$ space-time dimensions. In section \ref{S:Sound} we explain why, in a conformal theory, fourth sound should approach first sound at low temperatures
\begin{equation}
\label{E:fourthconformal}
	v_4^2 \xrightarrow[T \to 0]{} \frac{1}{d-1}
\end{equation}
where $d$ is the spacetime dimension.
Equation \eqref{E:fourthconformal} explains the asymptotic value of fourth sound found in \cite{Herzog:2008he} for dimension two operators but presents a puzzle for dimension one operators. In sections \ref{S:Setup} and \ref{S:Numerics} we use the Abelian Higgs model of \cite{Gubser:2008px,Hartnoll:2008vx} to study holographic superfluids with condensates of dimension $1/2 < \Delta  \leq 5$. We find that for $1/2 <\Delta \lsim 2$ conformal symmetry is apparently broken at low temperatures; the breaking of conformal symmetry introduces an extra scale which allows the asymptotic value of fourth sound to deviate from its expected conformal value. As we show in section \ref{S:Numerics} this extra scale can be tied to the divergence of the order parameter at low temperatures. Such a divergence, for dimension one operators, was first discussed in \cite{Hartnoll:2008vx}. In section \ref{S:uterm}  we adjust the matter action so that the scalar potential is bounded from below, enforcing conformal behavior at low temperatures. We summarize our findings in section \ref{S:Summary}.

\section{Sound modes in the Tisza-Landau  two-fluid model}
\label{S:Sound}
Hydrodynamics can be thought of as an effective theory, valid as long as the relevant fields vary slowly in space and time relative to some microscopic scale. In the absence of spontaneously broken symmetries, the hydrodynamic variables are the temperature, $T$, chemical potential, $\mu$, and fluid velocity, $u^m$ with $m=0, \ldots, d-1$.
In the presence of a spontaneously broken symmetry, the Goldstone boson $\varphi$ provides an extra degree of freedom. The hydrodynamic variables in such a  system are $T$, $\mu$, $u^m$ and $\xi^m = \partial^m \varphi$. 
In \cite{Son:2000ht} it was shown that, at the inviscid level, one may consistently construct a conserved energy momentum tensor and a conserved current out of these variables. Further, the resulting formulation of fluid dynamics coincides with the relativistic version of the Tisza-Landau two-fluid model of \cite{Khalatnikov, Carter:1993aq}. See also appendix A of \cite{Herzog:2008he}.

In the formulation of \cite{Son:2000ht} the energy momentum tensor, $T^{mn}$, takes the form
\begin{equation}
\label{E:Tmn}
	T^{mn} = \left(\epsilon + P\right)u^{m}u^{n} + P \eta^{mn}  + f^2 \xi^{m} \xi^{n}
\end{equation}
where the energy density, $\epsilon$, and the pressure, $P$, are related to the entropy, $s$, temperature, $T$, density of the normal phase, $\rhon$, and chemical potential, $\mu$, through
\begin{equation}
	\epsilon + P = T s + \mu \rhon\,.
\end{equation}
The density of the superfluid component is denoted $\rhos=f^2 \mu$ and the superfluid velocity is given by $v^m = \xi^m/\mu$. For the current one has
\begin{equation}
\label{E:current}
	J^{m} = \rhon u^m+f^2 \xi^m\,.
\end{equation}
In addition, the superfluid velocity satisfies a ``Josephson equation''
\begin{equation}
\label{E:Josephson}
	u^{m}\xi_m =  -\mu
\end{equation}
and we remember that it is generated by the gradient of the Goldstone boson $\varphi$,
\begin{equation}
\label{E:potential}
	\xi_m = \partial_m \varphi\,.
\end{equation}
In the absence of sources, the equations of motion are given by
\begin{align}
\label{E:emconservation}
	\partial_m T^{mn} &= 0 \\
\label{E:currentconservation}
	\partial_m J^m & = 0\,. 
\end{align}
It is often useful to decompose \eqref{E:emconservation} into a component orthogonal to the velocity field, and a component parallel to it. For the component parallel to the velocity field, we find from \eqref{E:potential}, \eqref{E:currentconservation} and
\begin{equation}
\label{E:dpval}
	d P = s dT + \rhon d\mu - f^2 \xi^m d \xi_m
\end{equation}
that
\begin{equation}
\label{E:entropy}
	0 = u_m \partial_n T^{mn} = \partial_m (s u^m)\,.
\end{equation}
This implies that the superfluid component does not carry entropy.

To study sound modes of the superfluid, we look at linear perturbations of the hydrodynamic variables around a static configuration of the fluid. In what follows, we will denote the unperturbed configuration with a subscript $0$ and the perturbed configuration with a $\delta$. Thus, $v^0 = 1$, $v^i = \delta v^i$, $u^0 = 1$, $u^i  = \delta u^i$, $\rhos =  {\rhos}_0 + \delta {\rhos}$ etc. The index $i$ runs over the spatial coordinates $i=1,\ldots,d-1$. At the linear level, \eqref{E:Josephson}, \eqref{E:potential}, \eqref{E:emconservation} and \eqref{E:currentconservation} take the form
\begin{subequations}
\label{E:lineoms}
\begin{align}
\label{E:linentropy}
	\partial_t \delta s + s_0 \partial_i \delta u^i &= 0 \\
\label{E:linstress}
	(\epsilon_0 + P_0) \partial_t \delta u^i +\mu_0 {\rhos}_0 \partial_t \delta v^i  & = - \partial_i \delta P \\
\label{E:lincurrent}
	\partial_t (\delta \rhos + \delta \rhon) +{ \rhon}_0 \partial_i \delta  u^i+{ \rhos}_0 \partial_i \delta  v^i & = 0\\
\label{E:linpotential}
	\mu_0 \partial_t \delta  v^i & = \partial_i \delta  \mu\,.
\end{align}
\end{subequations}
The first two equations correspond to the linearized version of \eqref{E:entropy} and the remaining components of \eqref{E:emconservation}, the third equation corresponds to the linearized current conservation equation \eqref{E:currentconservation} and the last equation follows from \eqref{E:Josephson} and \eqref{E:potential} .
It is possible to recast \eqref{E:lineoms} as wave equations for the total density $\rho = \rhos+\rhon$ and the entropy per particle $\sigma = s/\rho$, 
\begin{align}
\label{E:stt}
	\partial_t^2 \delta \sigma &= \frac{{\rhos}_0}{\rho_0}\frac{\sigma_0}{w_0}
		\left(\partial_i \partial^i \delta P - \frac{w_0 - {\rhos}_0\mu_0}{\mu_0} \partial_i \partial^i \delta\mu\right) \\ 
\label{E:rhott}
	\partial_t^2 \delta \rho &=  \frac{\rho_0}{w_0}\left(\frac{{\rhon}_0}{\rho_0} \partial_i \partial^i \delta P + \frac{T_0 {\rhos}_0 \sigma_0}{\mu_0} \partial_i \partial^i \delta \mu\right)
\end{align}
where we have denoted $w_0 = \epsilon_0+P_0$. If we now express the chemical potential and pressure in terms of the entropy per particle and density, we get a set of coupled wave equations. 

Consider the case where the superfluid phase is absent, $\rhos_0 = 0$. Then, the entropy does not oscillate and we get a density wave whose phase velocity squared is
\begin{equation}
\label{E:soundrhosiszero}
	v_1^2 = \frac{{\rhon}_0}{w_0}\left(\frac{\partial P}{\partial \rho}\right)_{\sigma}\,.
\end{equation}
This is the expected speed of sound in normal fluids. Once $\rhos \neq 0$ two coupled entropy-density waves are allowed. The phase velocities of the corresponding modes are the two roots of a second order polynomial. 
The bigger root reduces to \eqref{E:soundrhosiszero} when $\rhos$ is set to zero. The smaller root is the phase velocity of a new mode called second sound.
To measure second sound one can generate temperature waves in a tube filled with superfluid helium, and check when the wavelength of the mode matches the system size and generates a resonance. The first measurements of second sound in helium II are described in \cite{Lane}.

Now consider a setup where one allows only the superfluid component to flow. Such a situation can be achieved by scattering the normal, viscous, phase off stationary scatterers. In this case momentum is not conserved due to the force required to keep the scatterers in place. In such a setup we would set $\delta u^i = 0$ and remove the momentum conservation equation \eqref{E:linstress} from \eqref{E:lineoms}. Instead of \eqref{E:stt} and \eqref{E:rhott} we would have
\begin{align}
	\partial_t^2 \delta s &= 0 \\
\label{E:fourthrho}
	\partial_t^2 \delta \rho &= \frac{{\rhos}_0}{\mu_0} \partial_i \partial^i \mu.
\end{align}
Thus, we have one sound mode, similar to first sound since it excites a density wave, but with vanishing normal-component velocity. The existence of such a sound mode was first realized in \cite{Atkins} and is called fourth sound.
From \eqref{E:fourthrho} we find that the phase velocity of fourth sound is given by
\begin{equation}
\label{E:fourthsound}
	v_4^2 = \frac{{\rhos}_0}{\mu \left(\frac{\partial \rho}{\partial \mu}\right)_s}\,.
\end{equation}
Measurements of fourth sound can be carried out by allowing the superfluid to flow through a very narrow tube packed with fine powder which prevents the normal component from flowing. This is called a superleak, since it is only the super-component of the fluid that leaks into the tube. Transducers are then placed at the ends of the tube to receive and source the sound waves \cite{Shapiro}. 

In \cite{Herzog:2008he}, equation  \eqref{E:fourthsound} was used as an approximation to second sound at high temperature (or large entropy). At low temperatures most of the fluid will be in the superfluid phase so we can approximate $\rhos \sim \rho$.
Also, in a conformal fluid, where the only remaining scale at $T=0$ is $\mu$, we expect that $\rho \propto \mu^{d-1}$. Thus, for conformal fluids at low temperatures, fourth sound coincides with normal sound, 
\begin{equation}
\tag{\ref{E:fourthconformal}}
	v_4^2 \xrightarrow[T \to 0]{} \frac{1}{d-1}\,.
\end{equation}
This behavior is similar to that of fourth sound in ${}^4$He, where, as mentioned earlier, it interpolates between second sound close to the phase transition and first sound at low temperatures \cite{Atkins,Shapiro}.

\section{Setup}
\label{S:Setup}
Our construction of a holographic superfluid is very similar to that of \cite{Basu:2008bh,Herzog:2008he}. The bulk theory is the Abelian Higgs model studied in 
\cite{Gubser:2008px,Hartnoll:2008vx,Gubser:2008wz,Hartnoll:2008kx,Gubser:2008pf,Horowitz:2008bn}, whose action is given by
\begin{equation}
\label{E:Lagrangian}
    S =\frac{1}{2\kappa^2} \int \left( \mathcal{L}_{\rm EH} - \mathcal{L}_{\rm matter}\right) \sqrt{-g}d^4x\,
\end{equation}
with
\begin{equation}    
    \mathcal{L}_{\rm EH} = R - \Lambda\,,\quad 
    \mathcal{L}_{\rm matter} =  |\partial_{\mu} \Psi - i q A_{\mu} \Psi|^2 + V(|\Psi|^2) + \frac{1}{4} F_{\mu\nu}F^{\mu\nu}
\end{equation}
where $F_{\mu\nu} = \partial_{\mu}A_{\nu}-\partial_{\nu}A_{\mu}$ and our units are such that $\Lambda = -6$.  We will work in the probe approximation where the charge of the scalar field is very large, and the matter content of the theory decouples from gravity:
Setting $\Psi \to  \Psi/q$, $A_{\mu} \to A_{\mu}/q$, and scaling the coefficients in $V(|\Psi|^2)$ such that $V(|\Psi|^2) \to V(|\Psi|^2)/q^2$, implies $\mathcal{L}_{\rm matter} \to \mathcal{L}_{\rm matter}/q^2$. In the $q \to \infty$ limit $\mathcal{L}_{\rm matter}$ decouples from $\mathcal{L}_{\rm EH}$. As noted in \cite{Gubser:2008wv}, for large but finite $q$ this approximation is valid only as long as $\Psi$ and $\Phi$ are not too large. In principle, for any $T>0$, one may choose a large enough $q$ so that the approximation will be good enough.  Another way to think of the probe approximation is to consider a formal expansion of the full back-reacted geometry in inverse powers of $q$. The leading order matter solutions are $\mathcal{O}(q^{-1})$, while the leading order metric is $\mathcal{O}(q^0)$ and receives $\mathcal{O}(q^{-2})$ corrections.

Focusing first on $\mathcal{L}_{\rm EH}$, the solution to Einstein gravity in $3+1$ dimensions in the presence of a negative cosmological constant is given by the AdS-Schwarzschild black hole,
\begin{equation}
\label{E:AdSSS}
	g_{\mu\nu}dx^{\mu}dx^{\nu} = \frac{1}{z^2}\left(-f(z)dt^2+dx_i^2+\frac{dz^2}{f(z)}\right)\,,
\end{equation}
where 
\begin{equation}
	f(z)=1-\frac{z^3}{z_0^3}
\end{equation}
and $\mu=0,\ldots,3$.
The asymptotically AdS boundary is located at $z=0$ and the horizon is located at $z=z_0$.
The temperature of the black hole, $T$, is given by $z_0 = 3/4\pi T$.
Using $\Psi = \frac{1}{\sqrt{2}}\psi$ with $\psi$ real, $A_0 = \phi$ and $A_i = a_i$ (sometimes denoted $\vec{a}$), the equations of motion for the matter fields in the background \eqref{E:AdSSS} in the probe approximation are
\begin{subequations}
\label{E:phipsiaset}
\begin{align}
\label{E:psigeneral}
z^2 \partial_z \left( \frac{f}{z^2} \partial_z \psi \right) =& \left(a_i^2 - \frac{\phi^2}{f}+\frac{\partial_{\psi^{\star}}V(|\psi|^2)}{z^2}\right)\psi \\
\label{E:phigeneral}
\partial_z^2 \phi =& \frac{\psi^2 \phi}{z^2 f}
\\
\label{E:aigeneral}
\partial_z \left(f \partial_z a_i \right) =& \frac{\psi^2 a_i}{z^2}\,.
\end{align}
\end{subequations}
These configurations have been studied in \cite{Gubser:2008px,Hartnoll:2008vx,Basu:2008st,Horowitz:2008bn,Herzog:2008he} for 
\begin{equation}
\label{E:massterm}
	V(|\Psi|) = m^2 |\Psi|^2
\end{equation} 
and $m^2=-9/4,\,-2,$ and $0$. Using the standard AdS/CFT mapping, \cite{Gubser:1998bc,Witten:1998qj} the mass $m^2$ of the scalar field, $\psi$,  is related to the dimension $\Delta$ of its dual operator $\dualop$, through $m^2 = \Delta(3-\Delta)$. Unitarity implies that $\Delta \geq 1/2$.

The boundary conditions we apply to \eqref{E:phipsiaset} are that $\psi$ and $\vec{a}$ are finite at the horizon and that $\phi$ vanishes there. This fixes three of the six integration constants in the solution to \eqref{E:phipsiaset}. Near the asymptotically AdS boundary we require that $\phi(0)$ and $\vec{a}(0)$ are finite and that the $z^{3-\Delta}$ term in a near boundary series expansion of $\psi$ vanishes.

According to the AdS/CFT dictionary \cite{Gubser:1998bc,Witten:1998qj,Klebanov:1999tb},  the coefficient of the $z^{\Delta}$ term in a near boundary series expansion of $\psi$ is proportional to $\dualop$ while the coefficient of the $z^{3-\Delta}$ term is the source associated with $\dualop$. Our boundary conditions were chosen so that the source vanishes. The chemical potential $\mu$ and the charge density $\rho$ of the dual theory can be read off a near boundary expansion of $\phi$, $\phi = \mu - \rho z + \ldots$. Similarly, in the gauge we are using, the velocity field $\vec{\xi}$ and the spatial part of the current $\vec{J}$ are given by the expansion $\vec{a} = \vec{\xi} - \vec{J} z+\ldots$. See \cite{Herzog:2008he} for details. 

\section{Solving the equations of motion}
\label{S:Numerics}
\begin{figure}
\begin{center}
\scalebox{0.8}{\includegraphics{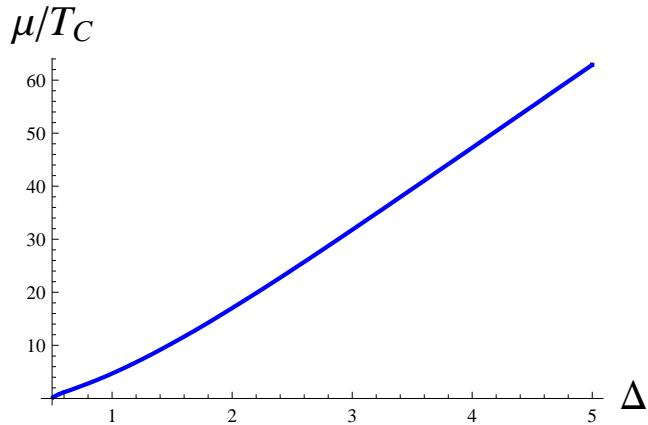}}
\caption{\label{F:Tc} The critical temperature $T_c$, where $\psi$ develops a zero mode, as a function of the dimension of the condensate $\Delta$ for $1/2< \Delta < 5$.}
\end{center}
\end{figure}

One solution to \eqref{E:phipsiaset}--\eqref{E:massterm} with the boundary conditions discussed in the previous section is
\begin{equation}
\label{E:normalsolution}
	\phi = \mu \left(1 - \frac{z}{z_0}\right) 
	\,,\quad
	\psi = 0
	\,,\quad
	\vec{a} = 0\,.
\end{equation}
Since $\dualop = 0$,  this solution describes a non-condensed phase. From the analysis of \cite{Gubser:2008px} we expect that at low enough temperatures other solutions to \eqref{E:phipsiaset}--\eqref{E:massterm} will exist, where the scalar field, $\psi$, is non vanishing.  Such solutions usually differ by the number of nodes of $\psi$. Since solutions with nodes are expected to be unstable we will focus on the solution where $\psi$ has no nodes. This will describe the condensed phase. At some critical temperature, $T_c$, the solution corresponding to the non-condensed phase and the solution corresponding to the condensed phase overlap: There exists a non trivial $\psi$ which solves \eqref{E:psigeneral} with $\phi$ and $\vec{a}$ as in \eqref{E:normalsolution}, which is determined only up to a multiplicative integration constant---a zero mode.\footnote{
In the case of symmetry breaking in a non-abelian setup in AdS${}_5$, an analytic expression for the zero mode was found in \cite{Basu:2008bh}. In \cite{Herzog:2009ci} it was shown that information about the phase transition can be obtained from this zero mode without the need to resort to numerics.}
Computing $T_c$ numerically is straightforward. It involves searching for a zero mode of the linear equation \eqref{E:psigeneral} with $\phi = \mu(1-z/z_0)$ and $\vec{a}=0$.
In figure \ref{F:Tc} we plot the dependence of the critical temperature on $\Delta$ as obtained from such a numerical analysis.
A similar plot, which goes beyond the probe approximation we are considering here, can be found in \cite{Denef:2009tp}. Thus, at least for $T<T_c$ two solutions to \eqref{E:phipsiaset} exist: a non-condensed solution given by \eqref{E:normalsolution} and a solution with a non vanishing expectation value for the condensate corresponding to the superfluid phase. If the transition is first order, the superfluid phase may also exist at $T>T_c$.

Following \cite{Herzog:2008he}, to see whether it is the superfluid phase or the non-condensed phase which is preferred, we need to compute the free energy $\Omega$ given by $\Omega = - T S_{\rm total}$ where $T$ is the temperature and $S_{\rm total}$ is the total renormalized matter action. This renormalized action is composed of two terms. The first is the on-shell matter action \eqref{E:Lagrangian},
\begin{equation}
\label{E:onshell}
	S_{\rm on-shell}  = \frac{V}{\kappa^2}
		\lim_{\epsilon \to 0} \left( \left(-\phi \partial_z \phi+f \vec{a} \cdot \partial_z \vec{a} + f\psi \partial_z \psi \right)\Big|_{\epsilon}^{z_0} + \int_{\epsilon}^{z_0}  \frac{\psi^2 \left(-\phi^2 f + |\vec{a}|^2 \right) }{z^2}dz \right)
\end{equation}
where $V$ is the volume of the transverse dimensions. The second contribution to $S_{\rm total}$ comes from holographically renormalizing the theory \cite{Bianchi:2001de,Bianchi:2001kw}, i.e., from adding boundary counterterms, $S_{\rm counter}$, to $S_{\rm on-shell}$ which render \eqref{E:onshell} finite. 
After inserting a near boundary expansion of $a_i$, $\phi$ and $\psi$ into \eqref{E:onshell}, we find that in the presence of non-vanishing sources for $\dualop$,  the only divergent contributions to $S_{\rm on-shell}$ come from the $z^{-2} \psi \partial_z \psi\Big|_{\epsilon}$ term on the right hand side of \eqref{E:onshell}.  Taking into account that the source terms for irrelevant operators should be treated as infinitesimal \cite{Witten:1998qj,Aharony:2005zr}, we find that only a mass counterterm is required to keep $S_{\rm on-shell}$ finite (meaning, a counterterm proportional to $\psi^2$). Since this counterterm contribution vanishes for the boundary conditions we are interested in, we are free to ignore it when computing $\Omega$. For all values of $\Delta$ we consider in this work, we found that the superfluid phase is preferred over the normal phase, and that the free energy is a smooth function of the temperature indicating a second order phase transition.

\begin{figure}
\begin{center}
\scalebox{0.8}{\includegraphics{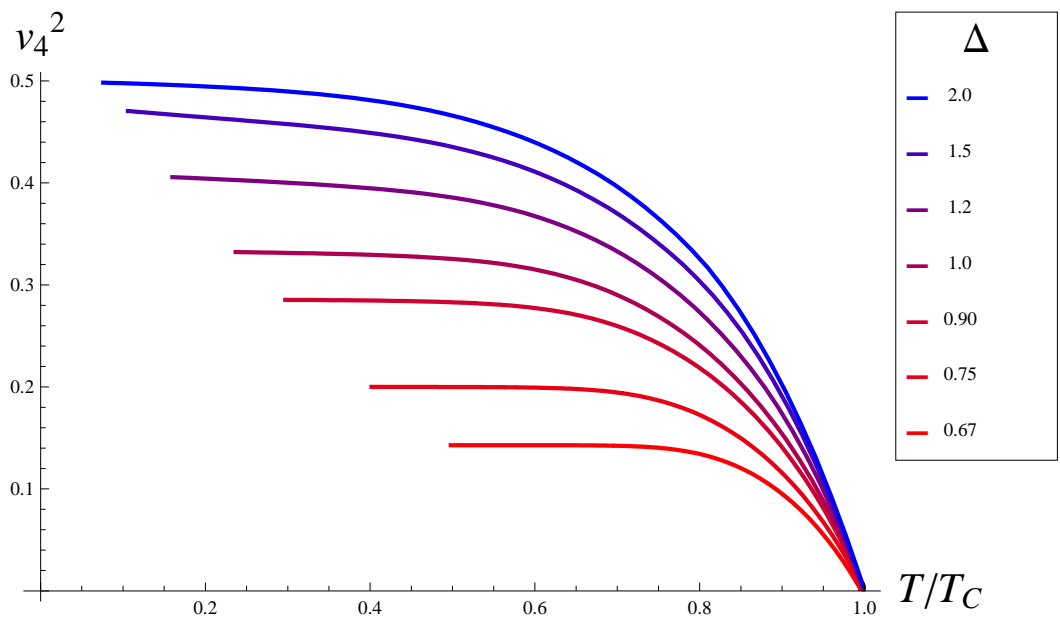}}
\caption{\label{F:4sound} (Color online) Fourth sound versus temperature for condensates of dimension $2/3 \leq \Delta \leq 2$ in the abelian Higgs model. The temperature is measured relative to the critical temperature $T_c$. The curves are color coded according to the values of $\Delta$.}
\end{center}
\end{figure}

\begin{figure}
\begin{center}
\scalebox{0.6}{\includegraphics{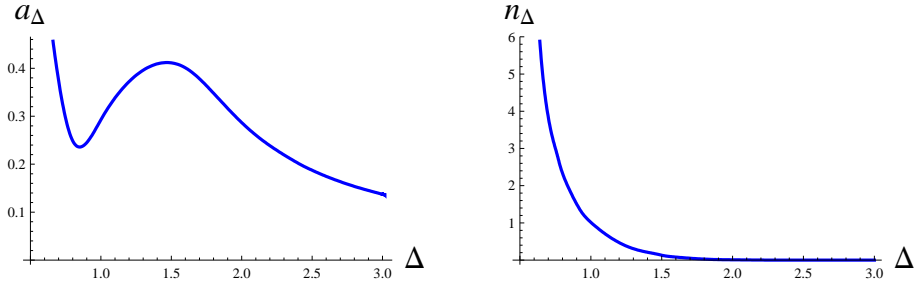}}
\caption{\label{F:rhobehavior} Values of $a_{\Delta}$ and $n_{\Delta}$, obtained by fitting the total charge density $\rho = a_{\Delta} \mu^2 \left(\frac{\mu}{T}\right)^{n_{\Delta}}$ to the numerics. The fit was carried out for data points for which $(\rho-\rhos)/\rho < 10^{-3}$ where $\rhos$ is the charge density of the superfluid phase. $\Delta$ is the dimension of the condensate.}
\end{center}
\end{figure}

We used a shooting algorithm to compute the solution to \eqref{E:phipsiaset}--\eqref{E:massterm} corresponding to the superfluid phase. Instead of shooting from the boundary, we fixed $\partial_z \phi(z_0)$ and $a(z_0)$ and searched for a value of $\psi(z_0)$ for which the near boundary $z^{3-\Delta}$ coefficient in a series expansion of $\psi$ vanishes. We did this once for $a(z_0)=0$ and then once more with a small enough value of $a(z_0)$ to obtain $\rhos =\mu  \lim_{\xi \to 0}|\vec{J}|/|\vec{\xi}|$ to a good approximation. This procedure was carried out for $\Delta$ between $13/25$ and $5$, with steps ranging from $1/10$ for $\Delta$ close to a half, to $1/2$ at larger $\Delta$. 
The resulting values of the speed of fourth sound, computed from \eqref{E:fourthsound}, for $\Delta \leq 2$ are shown in figure \ref{F:4sound}. 
For $\Delta > 2$, the curve for fourth sound closely follows that of $\Delta = 2$.
From figure \ref{F:4sound} we see that for $\Delta \lsim 2$ and the range of temperatures we are considering, the asymptotic values of $v_4^2$ deviate from the expected conformal behavior \eqref{E:fourthconformal}. The reason for this is the non conformal behavior of $\rho$ at low temperatures. Numerically, we find that for $(\rho-\rhos)/\rho < 10^{-3}$ one can approximate 
\begin{equation}
\label{E:rhobehavior}
	\rho = a_{\Delta} \mu^2 \left(\frac{\mu}{T}\right)^{n_{ \Delta}}
\end{equation}
where $a_{ \Delta}$ and $n_{\Delta}$ are plotted in figure \ref{F:rhobehavior}. Since $n_{ \Delta}>0$ for $\Delta\lsim 2$, the $T \to 0$ equation of state relating $\mu$ and $\rho$ deviates from the expected conformal one.

In some instances, it is difficult to decide from figure \ref{F:4sound} whether $v_4^2$ will reach its conformal value at $T=0$ or not; we find that for $3/2<\Delta<2$, the curve for $v_4^2$ seems to level off at a slower rate than that of other condensates. An extrapolation of the $v_4^2$ curve for $\Delta=3/2$ down to $T=0$ gives $v_4^2 \to 0.48$. Also, a least squares fit of the low temperature data to \eqref{E:rhobehavior} gives a much better result than fitting, say, a power law correction to a $\rho \propto \mu^2$ behavior. On the other hand, the $v_4^2$ curves for condensates of dimension $3/2 < \Delta < 2$ seem to posses an inflection point near $ 0.1 T_c$. 

\begin{figure}
\begin{center}
\scalebox{0.7}{\includegraphics{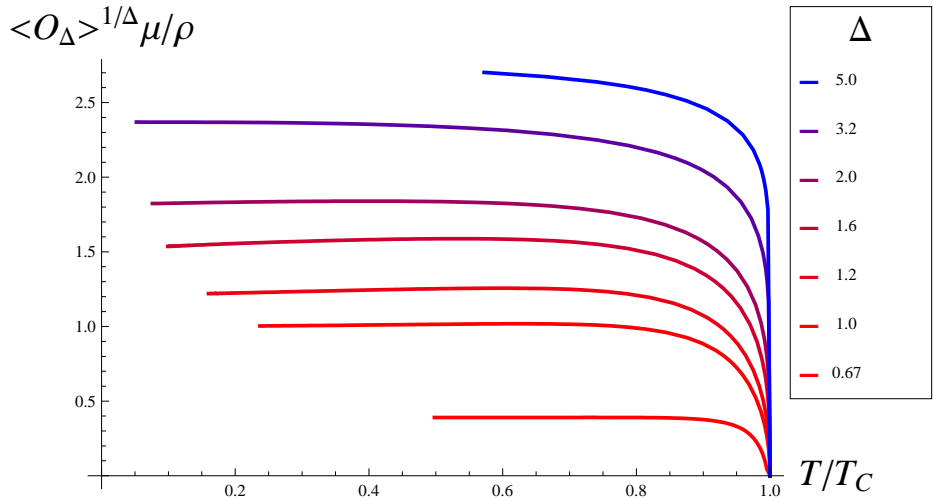}}
\caption{\label{F:Condensate} (Color online) The dimensionless ratio $\dualop^{1/\Delta}\mu/\rho$ versus temperature. At low temperatures this ratio approaches a constant, which implies via \eqref{E:rhobehavior} that $\dualop/\mu^{\Delta}$ diverges at low temperatures for small $\Delta$. The curves are color coded according to the value of $\Delta$.}
\end{center}
\end{figure}

In figure \ref{F:Condensate} we have plotted the dependence of the ratio $\dualop^{1/\Delta} \mu/\rho$ on $T/T_c$. Since $\dualop^{1/\Delta} \mu/\rho$ approaches a constant as $T \to 0$, this together with \eqref{E:rhobehavior} implies that $\dualop/\mu^{\Delta}$ diverges at low temperatures whenever $\Delta \lsim 2$. Thus, up to the numerical uncertainties discussed in the previous paragraph, the non conformal behavior of $\dualop$ is tied to the non-conformal behavior of $\rho$.

To better understand the low temperature limit of $\dualop$ and its relation to $\rho$, consider \eqref{E:phipsiaset} with $a=0$. Instead of the $z$ coordinate we will work with a rescaled radial coordinate $\zeta = z/\ell$ where $\ell$ is some length scale to be determined shortly. Defining also the rescaled variables $\zeta_0 = z_0/\ell$, $O= \dualop \ell^{\Delta}$, $R = \rho \ell^2$ and $M = \mu \ell$ and rescaling $\tilde{\phi} = \phi/\ell$, $\tilde{\psi} = \psi \zeta$ we can formally expand $\tilde{\phi}$ and $\tilde{\psi}$ in large $\zeta_0$. To leading order, we find
\begin{align}
\label{E:Phinearb}
	\partial_\zeta^2 \tilde{\phi} & =  \tilde{\phi} \tilde{\psi}^2\\
\label{E:Psinearb}
	\partial_\zeta^2 \tilde{\psi} & = \left(\frac{2-(3-\Delta)\Delta}{\zeta^2}-\tilde{\phi}^2\right)\tilde{\psi}.
\end{align}
As long as $ \tilde{\phi}^2 \tilde{\psi} \ll 1$ we may approximate  
\begin{equation}
\label{E:Psismallzeta}
	\tilde{\psi}(\zeta) = O \zeta^{\Delta-1}
\end{equation}
in which case the solution to \eqref{E:Phinearb} takes the form
\begin{equation}
\label{E:Phismallzeta}
	\tilde{\phi}(\zeta) = \left(\frac{O}{2\Delta}\right)^{1/{2\Delta}} \frac{2 M \zeta^{1/2}}{\Gamma\left(\frac{1}{2\Delta}\right)} \mathbf{K}_\frac{1}{2\Delta}\left(\frac{O \zeta^{\Delta}}{\Delta}\right)
\end{equation}
where $\mathbf{K}$ is a modified Bessel function of the second kind.
In \cite{Hartnoll:2008vx} it was observed that for $\Delta=1$, expression \eqref{E:Psismallzeta} for $\tilde{\psi}$ fits the low temperature numerical solution. 
The boundary conditions for \eqref{E:Phismallzeta} are that $\tilde{\phi}(0)=M$ and that $\tilde{\phi}$ does not increase exponentially in the deep interior. To fully justify the latter choice of boundary conditions we would need to match the large $\zeta$ expansion of the solution to \eqref{E:Phinearb} with a near horizon expansion of the solution to \eqref{E:phigeneral}. Such an analysis was carried out in a somewhat different context in \cite{Yarom:2007ap}. From \eqref{E:Phismallzeta} we can compute
\begin{equation}
\label{E:lowTasymptotics}
	\frac{\dualop^{1/\Delta} \mu}{\rho} = -\left(2 \Delta\right)^{1/\Delta}\frac{\Gamma\left(\frac{1}{2\Delta}\right)}{\Gamma\left(-\frac{1}{2\Delta}\right)}.
\end{equation}

\begin{figure}
\begin{center}
\scalebox{0.65}{\includegraphics{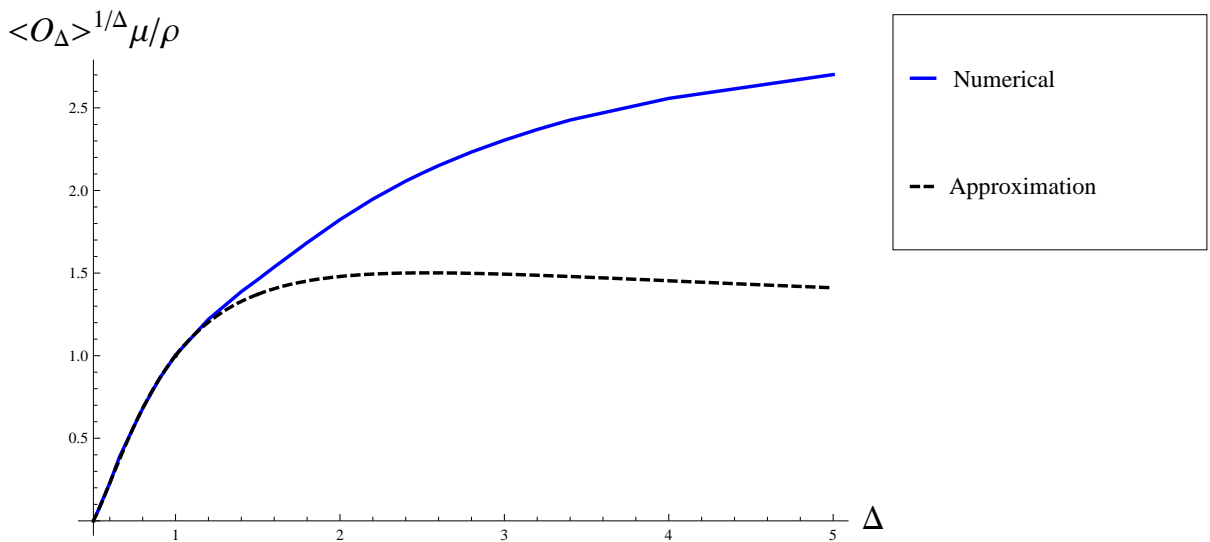}}
\caption{\label{F:fitlowT} (Color online) The asymptotic value of $\dualop^{1/\Delta}\mu/\rho$ at low temperatures, as a function of $\Delta$, the dimension of the scalar condensate. The blue curve shows the numerical result, displayed in figure \ref{F:Condensate}, and the dashed black curve shows the analytic approximation \eqref{E:lowTasymptotics}, valid for large $\dualop/\mu^{\Delta}$.}
\end{center}
\end{figure}

As stated earlier, \eqref{E:Psismallzeta} is valid only as long as the $\tilde{\phi}^2$ term in \eqref{E:Psismallzeta} is negligible. From \eqref{E:Phismallzeta} this implies that $M^2 O \ll 1$. Thus, we need to choose a scale $\ell$ for which both $z_0/\ell \gg 1$ and $\mu^2 \dualop \ell^{-\Delta-2} \ll 1$, and which would be compatible with our low temperature analysis, $z_0 \mu \gg 1$. If $\dualop/\mu^{\Delta}$ diverges at low temperatures (large $z_0 \mu$), we can use $\ell^{-\Delta} = \dualop$. 
In figure \ref{F:fitlowT} we plot our approximation \eqref{E:lowTasymptotics} versus the numerical value of $\dualop^{1/\Delta}\mu/\rho$ for $1/2 < \Delta < 5$. 

\section{Bounding the scalar potential}
\label{S:uterm}

It is tempting to think of \eqref{E:Phinearb} and \eqref{E:Psinearb} as the zero temperature description of the condensate. Recall however, that we're working in the probe approximation where the matter fields are small enough so that they don't distort the metric. In practice, once $\psi$ and $\phi$ become too large the geometry will back-react on the matter fields. This will certainly have a significant effect whenever $\Delta \lsim 2$, as indicated by the divergent behavior of $\dualop/\mu^{\Delta}$, or by the non conformal behavior of fourth sound at low temperatures. One way to ensure a conformally invariant state at low temperatures, valid in the probe approximation, is to consider a scalar potential, $V$, which has a minimum at a finite value of $|\Psi|$. This is exactly the situation considered in \cite{Gubser:2008wz}, where the equations of motion of the full gravity-matter action of \eqref{E:Lagrangian} were studied for a scalar potential of the form
\begin{equation}
\label{E:Vterm}
	V(|\psi|) = m^2 |\psi|^2 +\frac{1}{2} u^2 |\psi|^4\,
\end{equation} 
with $m^2 = -2$.
In the presence of the potential \eqref{E:Vterm} there exists an empty AdS solution to the equations of motion whenever $\psi$ is at an extremum of $V$, and $\phi$ and $a_i$ vanish. As shown in \cite{Gubser:2008wz}, this allows for a condensed phase where $\psi$ flows from $\psi=0$ near the boundary to $\psi = \sqrt{2}/u$ in the infrared. Since the geometry is AdS${}_4$ both in the ultraviolet and in the deep interior, one can imagine a limit where it is approximately AdS${}_4$ throughout, ensuring the validity of the probe approximation near $T=0$. In this case we expect that close to $T=0$, $\dualop \propto \mu^{\Delta}$ and $v_4^2 \to 1/2$. This is indeed the case, as we show below.

\begin{figure}
\begin{center}
\scalebox{0.8}{\includegraphics{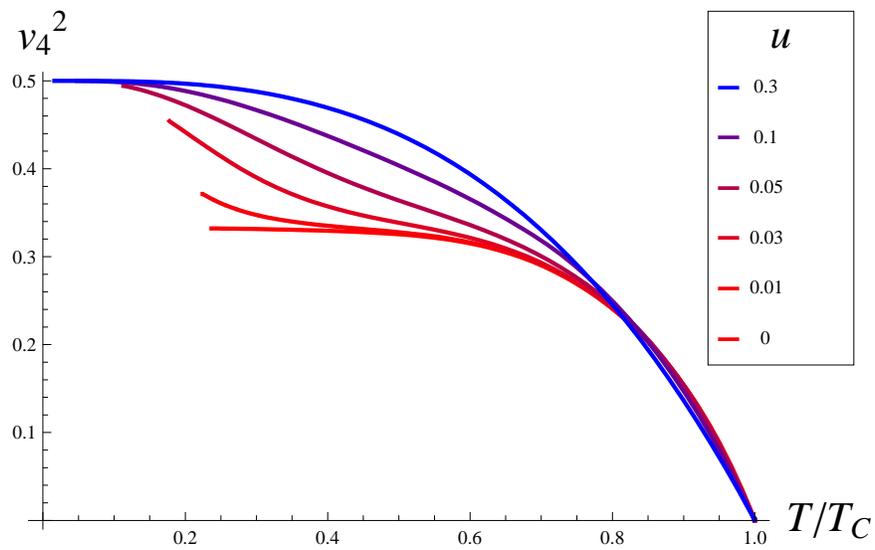}}
\caption{\label{F:fourthsoundu} (Color online) Fourth sound versus temperature for a dimension one condensate and a potential as in \eqref{E:Vterm}. The curves are color coded according to the values of $u$. As $u$ increases, lower temperatures become  numerically accessible.}
\end{center}
\end{figure}

In the probe approximation, the equations of motion for the gauge field and scalar field with a scalar potential as in \eqref{E:Vterm} can be read off of \eqref{E:phipsiaset}. The boundary conditions we used and the numerical procedure for solving the equations of motion are similar to those discussed in sections \ref{S:Setup} and \ref{S:Numerics}.
We obtained numerical solutions 
for $\Delta = 1$ and $\Delta = 2$ and $10^{-3} \leq u^2 \leq 1/3$ in steps as small as $10^{-2}$.
The resulting value of fourth sound for dimension one operators is plotted in figure \ref{F:fourthsoundu}.
We find that, as expected, once $u$ is non vanishing, fourth sound tends to reach its conformal value, $v_4^2 = 1/2$ at low temperatures. At intermediate temperatures, and for relatively small $u$, fourth sound seems to pass a transition region where $v_4^2$ is approximately $1/3$ as was the case for $u=0$. The fact that $v_4^2$ tends to $1/2$ at low temperatures indicates that conformal invariance is retained. Indeed, for $2/15 < u^2 < 1/3$ we find that 
\begin{equation}
\label{E:rhowithu}
	\rho = a u^{-n} \mu^2
\end{equation}
with $a=0.69$ and $n=1.07$ to a fairly good approximation. For dimension two operators and the same range of $u$, we obtain \eqref{E:rhowithu} with $a=0.28$ and $n=3 \times 10^{-3}$.

Computing $\dualop$ for $\Delta=1,\,2$ we find that, unsurprisingly, once $u \neq 0$, $\dualop^{1/\Delta} \mu/\rho$ asymptotes to a finite value at small $T$. An analytic expression for the low temperatures dependence of $\dualop$ on $u$, $\mu$, $\rho$, $\vec{\xi}$ and $\vec{J}$ can be obtained as follows. Consider \eqref{E:phipsiaset} with a potential $V$ as in \eqref{E:Vterm}. Defining $\hat \psi = \psi z $, and setting $z_0\mu \gg 1$, the low temperature equations of motion take the form
\begin{subequations}
\label{E:EOMu}
\begin{align}
	\partial_z^2 \hat{\psi} &= -{\phi}^2 \hat{\psi} + |\vec{a}|^2 \hat{\psi} + u^2 \hat{\psi}^3 \\
	\partial_z^2 {\phi} & =  {\phi} \hat{\psi}^2 \\
	\partial_z^2 \vec{a} & =  \vec{a} \hat{\psi}^2\,.
\end{align}
\end{subequations}
Since $z$ does not appear explicitly in \eqref{E:EOMu}, we can construct the conserved quantity,
\begin{equation}
\label{E:Hamiltonian}
	\mathcal{H} = \frac{1}{2}\left((\partial_z \hat{\psi})^{2}+|\partial_z \vec{a}|^{2}-(\partial_z \phi)^{2}\right)
		+\frac{1}{2}\left(\phi^2-|\vec{a}|^2-\frac{1}{2}u^2 \hat{\psi}^2\right)\hat{\psi}^2 \, .
\end{equation}
Deep in the infrared, at very large $z$, we expect the scalar field to be at the minimum of the potential \eqref{E:Vterm}, $\psi = \sqrt{2}/{u}$, and the other fields to vanish, $\vec{a} = 0$ and $\phi = 0$. With this information, evaluating \eqref{E:Hamiltonian} at large $z$ we find
\begin{equation}
\label{E:His0}
	\mathcal{H}=0\,.
\end{equation} 
Using \eqref{E:His0} and the near boundary expansion of $\psi$, $\phi$ and $\vec{a}$ in \eqref{E:Hamiltonian}, we obtain
\begin{equation}
\label{E:conserved2}
	\langle O_2 \rangle^2 = \rho^2 - |\vec{J}|^2
\end{equation}
for dimension two operators, and
\begin{equation}
\label{E:conserved1}
	\langle O_1 \rangle^2 = u^{-2} \left( (\mu^2-|\vec{\xi}|^2) \pm \sqrt{\left(\mu^2-|\vec{\xi}|^2\right)^2 - 2 u^2 \left(\rho^2-|\vec{J}|^2\right)}\right)
\end{equation}
for dimension one operators. If the superfluid velocity vanishes, $\xi=J=0$, we find that in the $u \to 0$ limit \eqref{E:conserved2} reduces to 
\begin{equation}
\label{E:conserved2u}
	\frac{\langle O_2 \rangle}{\rho} = 1
\end{equation}
while the `-' branch of \eqref{E:conserved1} reduces to
\begin{equation}
\label{E:conserved1u}
	\frac{\langle O_1 \rangle \mu}{\rho} = 1
\end{equation}
which coincides with \eqref{E:lowTasymptotics} for $\Delta=1$. We've checked that \eqref{E:conserved2}, \eqref{E:conserved1}, \eqref{E:conserved2u} and \eqref{E:conserved1u} agree with our numerical results and with those in the literature \cite{Herzog:2008he}.

\section{Summary}
\label{S:Summary}

The results of our analysis can be summarized by figures \ref{F:4sound}, \ref{F:rhobehavior}, \ref{F:Condensate} and \ref{F:fourthsoundu}: In figure \ref{F:4sound} we see that in $d=2+1$ dimensions fourth sound deviates from its expected conformal behavior at low temperatures  for condensates of dimension $\Delta\lsim 2$, due to an anomalous scaling of the charge density (figure \ref{F:rhobehavior}). Figure \ref{F:Condensate} together with figure \ref{F:rhobehavior} relate the anomalous scaling of the charge density to the divergent behavior of $\dualop/\mu^{\Delta}$ at low temperatures. Finally, in figure \ref{F:fourthsoundu} we see how one can enforce conformal behavior at low temperatures by ensuring an asymptotically AdS solution in the deep interior. The relation we've found between the charge density and the value of the condensate seems quite robust: one may summarize it by saying that if $\dualop/\mu^{\Delta}$ is finite at low temperatures then fourth sound will asymptote to its conformal value. This behavior is in agreement with the $3+1$ dimensional p-wave superfluid studied in \cite{Herzog:2009ci}.

Apart from the asymptotic behavior of $\dualop$, described at the end of sections \ref{S:Numerics} and \ref{S:uterm}, our results were numerical. As such, we were unable to obtain a solution at arbitrarily low temperatures. For $u=0$ and condensates of dimension $\Delta\lsim 2$ it is difficult to study the low temperature solutions due to the divergent behavior of $\rho$ and $\dualop$. For large values of $\Delta$, keeping track of (at least) the first $2 \Delta -3 $ terms in a near boundary series expansion of $\psi$ becomes challenging as $z_0\mu$ becomes large. Unfortunately, this numerical issue makes it hard to find an exact critical dimension, $\Delta_c$, below which the the scaling dimension of the condensate becomes anomalous at low temperatures. To obtain a better estimate of $\Delta_c$, we need to go down to temperatures of at least half the current value obtained. With our current algorithm, probing such low temperatures would require significantly more precision than the 40 digits of precision we have worked with.

Experimental results for fourth sound in ${}^4$He \cite{Shapiro} closely follow the theoretical curve predicted in \cite{Atkins}. It is satisfying that for condensates with $\Delta \gsim 2$ the dependance of fourth sound of holographic superfluids on the temperature  is very similar to that of helium II. With such a qualitatively good fit to fourth sound, it would be interesting to compute second sound by solving the equations of motion for the full back-reacted geometry.


\section*{Acknowledgments}
I'd like to thank  F. Benini, S. Gubser, C. Herzog, A. Nellore, S. Pufu and F. Rocha  for many interesting discussions and for help with the numerics. This work is supported by  the Department of Energy under Grant No. DE-FG02-91ER40671.

\bibliographystyle{JHEP}
\bibliography{Teq0}

\providecommand{\href}[2]{#2}\begingroup\raggedright\begin{thebibliography}{10}

\bibitem{Tisza}
T.~L., {\it {Transport phenomena in helium {II}}},  {\em Nature} {\bf 141}
  (1938) 913.

\bibitem{Landau}
L.~D. Landau, {\it {The theory of superfluidity of helium II}},  {\em J. Phys.
  USSR} {\bf 5} (1941) 71.

\bibitem{Maldacena:1997re}
J.~M. Maldacena, {\it {The large N limit of superconformal field theories and
  supergravity}},  {\em Adv. Theor. Math. Phys.} {\bf 2} (1998) 231--252,
  [\href{http://xxx.lanl.gov/abs/hep-th/9711200}{{\tt hep-th/9711200}}].

\bibitem{Gubser:1998bc}
S.~S. Gubser, I.~R. Klebanov, and A.~M. Polyakov, {\it {Gauge theory
  correlators from non-critical string theory}},  {\em Phys. Lett.} {\bf B428}
  (1998) 105--114, [\href{http://xxx.lanl.gov/abs/hep-th/9802109}{{\tt
  hep-th/9802109}}].

\bibitem{Witten:1998qj}
E.~Witten, {\it {Anti-de Sitter space and holography}},  {\em Adv. Theor. Math.
  Phys.} {\bf 2} (1998) 253--291,
  [\href{http://xxx.lanl.gov/abs/hep-th/9802150}{{\tt hep-th/9802150}}].

\bibitem{Policastro:2001yc}
G.~Policastro, D.~T. Son, and A.~O. Starinets, {\it {The shear viscosity of
  strongly coupled N = 4 supersymmetric Yang-Mills plasma}},  {\em Phys. Rev.
  Lett.} {\bf 87} (2001) 081601,
  [\href{http://xxx.lanl.gov/abs/hep-th/0104066}{{\tt hep-th/0104066}}].

\bibitem{Bhattacharyya:2008jc}
S.~Bhattacharyya, V.~E. Hubeny, S.~Minwalla, and M.~Rangamani, {\it {Nonlinear
  Fluid Dynamics from Gravity}},  {\em JHEP} {\bf 02} (2008) 045,
  [\href{http://xxx.lanl.gov/abs/0712.2456}{{\tt arXiv:0712.2456}}].

\bibitem{Gubser:2008px}
S.~S. Gubser, {\it {Breaking an Abelian gauge symmetry near a black hole
  horizon}},  \href{http://xxx.lanl.gov/abs/0801.2977}{{\tt arXiv:0801.2977}}.

\bibitem{Hartnoll:2008vx}
S.~A. Hartnoll, C.~P. Herzog, and G.~T. Horowitz, {\it {Building a Holographic
  Superconductor}},  {\em Phys. Rev. Lett.} {\bf 101} (2008) 031601,
  [\href{http://xxx.lanl.gov/abs/0803.3295}{{\tt arXiv:0803.3295}}].

\bibitem{Gubser:2008zu}
S.~S. Gubser, {\it {Colorful horizons with charge in anti-de Sitter space}},
  {\em Phys. Rev. Lett.} {\bf 101} (2008) 191601,
  [\href{http://xxx.lanl.gov/abs/0803.3483}{{\tt arXiv:0803.3483}}].

\bibitem{Gubser:2008wv}
S.~S. Gubser and S.~S. Pufu, {\it {The gravity dual of a p-wave
  superconductor}},  {\em JHEP} {\bf 11} (2008) 033,
  [\href{http://xxx.lanl.gov/abs/0805.2960}{{\tt arXiv:0805.2960}}].

\bibitem{Ammon:2008fc}
M.~Ammon, J.~Erdmenger, M.~Kaminski, and P.~Kerner, {\it {Superconductivity
  from gauge/gravity duality with flavor}},
  \href{http://xxx.lanl.gov/abs/0810.2316}{{\tt arXiv:0810.2316}}.

\bibitem{Basu:2008bh}
P.~Basu, J.~He, A.~Mukherjee, and H.-H. Shieh, {\it {Superconductivity from
  D3/D7: Holographic Pion Superfluid}},
  \href{http://xxx.lanl.gov/abs/0810.3970}{{\tt arXiv:0810.3970}}.

\bibitem{Basu:2008st}
P.~Basu, A.~Mukherjee, and H.-H. Shieh, {\it {Supercurrent: Vector Hair for an
  AdS Black Hole}},  \href{http://xxx.lanl.gov/abs/0809.4494}{{\tt
  arXiv:0809.4494}}.

\bibitem{Herzog:2008he}
C.~P. Herzog, P.~K. Kovtun, and D.~T. Son, {\it {Holographic model of
  superfluidity}},  \href{http://xxx.lanl.gov/abs/0809.4870}{{\tt
  arXiv:0809.4870}}.

\bibitem{Herzog:2009ci}
C.~P. Herzog and S.~S. Pufu, {\it {The Second Sound of SU(2)}},
  \href{http://xxx.lanl.gov/abs/0902.0409}{{\tt arXiv:0902.0409}}.

\bibitem{Atkins}
K.~R. Atkins, {\it Third and fourth sound in liquid helium ii},  {\em Phys.
  Rev.} {\bf 113} (Feb, 1959) 962--965.

\bibitem{PhysRev.73.608}
J.~R. Pellam, {\it Wave transmission and reflection phenomena in liquid helium
  {II}},  {\em Phys. Rev.} {\bf 73} (Mar, 1948) 608--617.

\bibitem{Son:2000ht}
D.~T. Son, {\it {Hydrodynamics of relativisic systems with broken continuous
  symmetries}},  {\em Int. J. Mod. Phys.} {\bf A16S1C} (2001) 1284--1286,
  [\href{http://xxx.lanl.gov/abs/hep-ph/0011246}{{\tt hep-ph/0011246}}].

\bibitem{Khalatnikov}
I.~M. Khalatnikov and V.~V. Lebedev, {\it {Relativistic hydrodynamics of a
  superfluid liquid}},  {\em Phys. Lett.} {\bf 91 A} (1982).

\bibitem{Carter:1993aq}
B.~Carter and I.~M. Khalatnikov, {\it {Canonically covariant formulation of
  Landau's Newtonian superfluid dynamics}},  {\em Rev. Math. Phys.} {\bf 6}
  (1994) 277--304.

\bibitem{Lane}
C.~Lane, A.~Fairbank, and W.~M. Fairbank, {\it {Second sound in liquid helium
  II}},  {\em Phys. Rev.} {\bf 71} (1947) 600.

\bibitem{Shapiro}
K.~A. Shapiro and I.~Rudnick, {\it Experimental determination of fourth sound
  velocity in helium ii},  {\em 137} (1965).

\bibitem{Gubser:2008wz}
S.~S. Gubser and F.~D. Rocha, {\it {The gravity dual to a quantum critical
  point with spontaneous symmetry breaking}},
  \href{http://xxx.lanl.gov/abs/0807.1737}{{\tt arXiv:0807.1737}}.

\bibitem{Hartnoll:2008kx}
S.~A. Hartnoll, C.~P. Herzog, and G.~T. Horowitz, {\it {Holographic
  Superconductors}},  \href{http://xxx.lanl.gov/abs/0810.1563}{{\tt
  arXiv:0810.1563}}.

\bibitem{Gubser:2008pf}
S.~S. Gubser and A.~Nellore, {\it {Low-temperature behavior of the Abelian
  Higgs model in anti-de Sitter space}},
  \href{http://xxx.lanl.gov/abs/0810.4554}{{\tt arXiv:0810.4554}}.

\bibitem{Horowitz:2008bn}
G.~T. Horowitz and M.~M. Roberts, {\it {Holographic Superconductors with
  Various Condensates}},  {\em Phys. Rev.} {\bf D78} (2008) 126008,
  [\href{http://xxx.lanl.gov/abs/0810.1077}{{\tt arXiv:0810.1077}}].

\bibitem{Klebanov:1999tb}
I.~R. Klebanov and E.~Witten, {\it {AdS/CFT correspondence and symmetry
  breaking}},  {\em Nucl. Phys.} {\bf B556} (1999) 89--114,
  [\href{http://xxx.lanl.gov/abs/hep-th/9905104}{{\tt hep-th/9905104}}].

\bibitem{Denef:2009tp}
F.~Denef and S.~A. Hartnoll, {\it {Landscape of superconducting membranes}},
  \href{http://xxx.lanl.gov/abs/0901.1160}{{\tt arXiv:0901.1160}}.

\bibitem{Bianchi:2001de}
M.~Bianchi, D.~Z. Freedman, and K.~Skenderis, {\it {How to go with an RG
  flow}},  {\em JHEP} {\bf 08} (2001) 041,
  [\href{http://xxx.lanl.gov/abs/hep-th/0105276}{{\tt hep-th/0105276}}].

\bibitem{Bianchi:2001kw}
M.~Bianchi, D.~Z. Freedman, and K.~Skenderis, {\it {Holographic
  Renormalization}},  {\em Nucl. Phys.} {\bf B631} (2002) 159--194,
  [\href{http://xxx.lanl.gov/abs/hep-th/0112119}{{\tt hep-th/0112119}}].

\bibitem{Aharony:2005zr}
O.~Aharony, A.~Buchel, and A.~Yarom, {\it {Holographic renormalization of
  cascading gauge theories}},  {\em Phys. Rev.} {\bf D72} (2005) 066003,
  [\href{http://xxx.lanl.gov/abs/hep-th/0506002}{{\tt hep-th/0506002}}].

\bibitem{Yarom:2007ap}
A.~Yarom, {\it {The high momentum behavior of a quark wake}},  {\em Phys. Rev.}
  {\bf D75} (2007) 125010, [\href{http://xxx.lanl.gov/abs/hep-th/0702164}{{\tt
  hep-th/0702164}}].

\end{thebibliography}\endgroup

\end{document}